\documentstyle[elsart12,osa,epsf,eqsecnum,epsfig,graphics,cite]{revtex}

\def\lsim{\raise0.3ex\hbox{$<$\kern-0.75em\raise-1.1ex\hbox{$\sim$}}}
\def\gsim{\raise0.3ex\hbox{$>$\kern-0.75em\raise-1.1ex\hbox{$\sim$}}}

\def\beq{\begin{equation}}
\def\eeq{\end{equation}}
\def\bea{\begin{eqnarray}}
\def\eea{\end{eqnarray}}

\renewcommand{\lsim}{~{\buildrel < \over {_\sim}}~}
\renewcommand{\gsim}{~{\buildrel > \over {_\sim}}~}

\renewcommand{\d}{{\rm d}}

\begin{document}


\voffset1.5cm


\title{\Large\bf How sensitive are high-$p_T$ electron spectra at RHIC 
to heavy quark energy loss?}
\author{N\'estor Armesto$^{\rm 1}$, Matteo Cacciari$^{\rm 2}$, 
Andrea Dainese$^{\rm 3}$,
Carlos A.~Salgado$^{\rm 4}$ and Urs Achim Wiedemann$^{\rm 4,5}$}

\address{
$^{\rm 1}$ Departamento de F\'{\i}sica de Part\'{\i}culas and IGFAE,
Universidade de Santiago de Compostela, 
Spain\\
$^{\rm 2}$ LPTHE, Universit\'e Pierre et Marie Curie (Paris 6), France\\
$^{\rm 3}$ Universit\`a degli Studi di Padova and INFN, Padova, Italy\\
$^{\rm 4}$ Department of Physics, CERN, Theory Division,
CH-1211 Gen\`eve 23, Switzerland\\
$^{\rm 5}$Department of Physics and Astronomy, University of Stony Brook, NY 11794, USA
}
\date{\today}
\maketitle

\begin{abstract} 
In nucleus--nucleus collisions, high-$p_T$ electron spectra depend on the medium modified
fragmentation of their massive quark parents, thus giving novel access to the predicted mass
hierarchy of parton energy loss. Here we calculate these spectra in a model, which
supplements the perturbative QCD factorization formalism with parton energy loss.
In general, we find - within large errors - rough agreement between theory and 
data on the single inclusive electron spectrum in pp, its nuclear modification factor 
$R_{AA}^e$, and its azimuthal anisotropy $v_2^e$. However, the nuclear modification factor 
depends on the relative contribution of charm and
bottom production, which we find to be affected by large perturbative uncertainties.
In order for electron measurements to provide a significantly more stringent test of the expected mass 
hierarchy,  one must then disentangle the  
$b$- and $c$-decay contributions, for instance by reconstructing the 
displaced decay vertices. 
\end{abstract}

\section{Introduction}
\label{sec1}

The semi-leptonic decays of charmed and beauty mesons dominate the high-$p_T$ 
electron spectrum in $\sqrt{s_{NN}} = 200$~GeV hadronic collisions at the
Brookhaven Relativistic Heavy Ion Collider (RHIC) up to $p_T \approx 20$~GeV,
where the Drell--Yan contribution starts to become significant. At the
30 times higher center of mass energies of the CERN Large Hadron Collider (LHC),
heavy quark decays are expected to dominate the electron spectrum up to
$p_T \approx 30$ -- $35$~GeV, where $W$-decay contributions take over. 
In relativistic nucleus--nucleus collisions, such electron spectra are of great interest,
since they are expected to give access to the medium modified fragmentation of
their heavy quark parents~\cite{Dokshitzer:2001zm,Armesto:2003jh,Zhang:2003wk,Djordjevic:2003zk}. Models of strongly medium-enhanced  fragmentation 
of light quarks and gluons~\cite{Gyulassy:2003mc,Jacobs:2004qv,Wang:2003aw,Dainese:2004te,Eskola:2004cr} are favored by RHIC data, which show
a strong suppression of high-$p_T$ hadron spectra and hadron-triggered
correlation functions~\cite{Adcox:2004mh,Back:2004je,Arsene:2004fa,Adams:2005dq}.
Establishing the dependence of this suppression on the mass and color charge 
of the parent parton provides a novel opportunity to further test the 
microscopic picture  conjectured to underlie medium-induced
high-$p_T$ hadron suppression~\cite{Armesto:2005iq}. This may also help to further constrain 
information about the density and collective dynamics of the QCD matter produced in 
heavy ion collisions, for which parton energy loss is considered to be a sensitive probe. 

Preliminary RHIC data on single inclusive electrons in Au--Au 
collisions~\cite{Adler:2005xv,Bielcik:2005wu} extend 
over a wider transverse momentum range, up to $p_T^e \sim 10$~GeV,
than previously published results~\cite{Adcox:2002cg}. 
The purpose of this letter is to assess to what extent  these spectra
and their azimuthal asymmetry~\cite{Sakai:2005qn,Laue:2004tf} are in agreement 
with our current understanding of medium-induced parton energy loss, and 
how they can help us to access in more detail the microscopic 
mechanisms underlying high-$p_T$ hadron suppression. To
this end, we determine the sensitivity of the nuclear modification factor for single 
electrons $R_{AA}^{e}(p_T)$ on uncertainties in the $b$- and $c$-benchmark 
cross sections, we quantify the possible contamination of $R_{AA}^{e}(p_T)$
from other hard processes (namely Drell--Yan), and we study its dependence on those properties 
of the nuclear matter, which are expected to induce the energy 
degradation of hard parent partons.

\section{Uncertainties of the proton--proton benchmark}

\begin{figure}[t]
\begin{center}
\includegraphics[width=10.0cm]{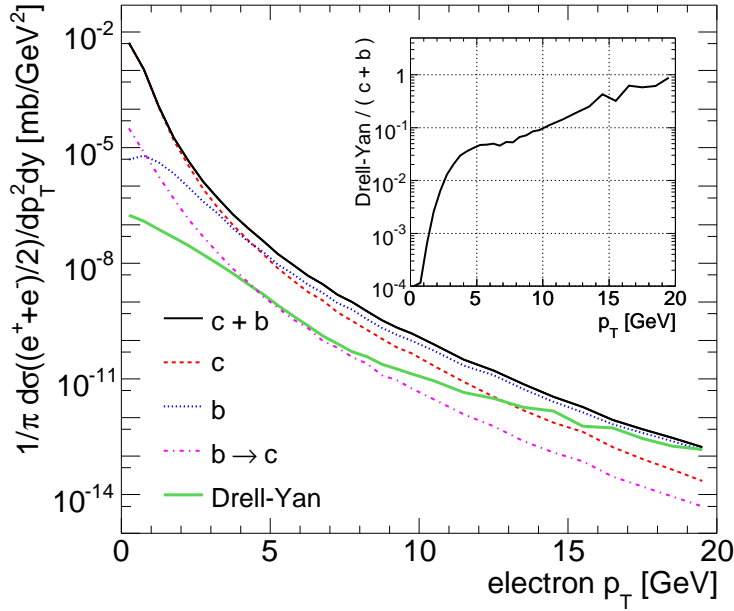}
\end{center}
\vspace{0.5cm}
\caption{(Color online) Contributions from semi-leptonic heavy quark decays and from
Drell--Yan pairs to the single inclusive electron spectrum in proton-proton collisions at 
$\sqrt{s} = 200$~GeV, calculated with NLO accuracy.
}\label{fig1}
\end{figure}

Heavy quark production in proton--proton collisions can be calculated in perturbative QCD via the 
collinear factorization 
approach~\cite{Nason:1987xz,Nason:1989zy,Beenakker:1990ma}. 
In Fig.~\ref{fig1}, we show the resulting electron spectrum for $\sqrt{s_{NN}}=200$~GeV
pp collisions at RHIC.  This spectrum was obtained~\cite{Cacciari:2005rk} by complementing
a fixed-order plus next-to-leading-log-resummed
(FONLL~\cite{Cacciari:1998it,Cacciari:2001td}) heavy quark $p_T$ distribution with proper fragmentation functions, 
describing the hadronization into heavy hadrons, and with the subsequent 
decay of the heavy hadrons into electrons.
 The Drell--Yan contribution, also shown in  Fig.~\ref{fig1}, is obtained using
PYTHIA~\cite{Sjostrand:2001yu}, with a Drell--Yan cross section rescaled such  that the
$e^+\, e^-$ invariant mass distribution matches the one given by the NLO calculation in
Ref.~\cite{Gavin:1995ch}. The curves in Fig.~\ref{fig1} reflect central values of a calculation, in
which the Drell--Yan contribution to the single electron spectrum is about
10\% at $p_T^e=10$~GeV. 
This implies that neglecting the Drell--Yan contribution in computations of  the nuclear
modification factor of single electrons, one reduces $R_{AA}^e$ by up to $0.1$.   
However, the theoretical uncertainties  are large. At smaller transverse momenta $p_T^e$, 
on which we focus in this
paper, Drell--Yan becomes even smaller (see Fig.~\ref{fig1}), and will be neglected in
what follows.

\begin{figure}[t]
\begin{center}
\includegraphics[width=8.3cm,angle=-90]{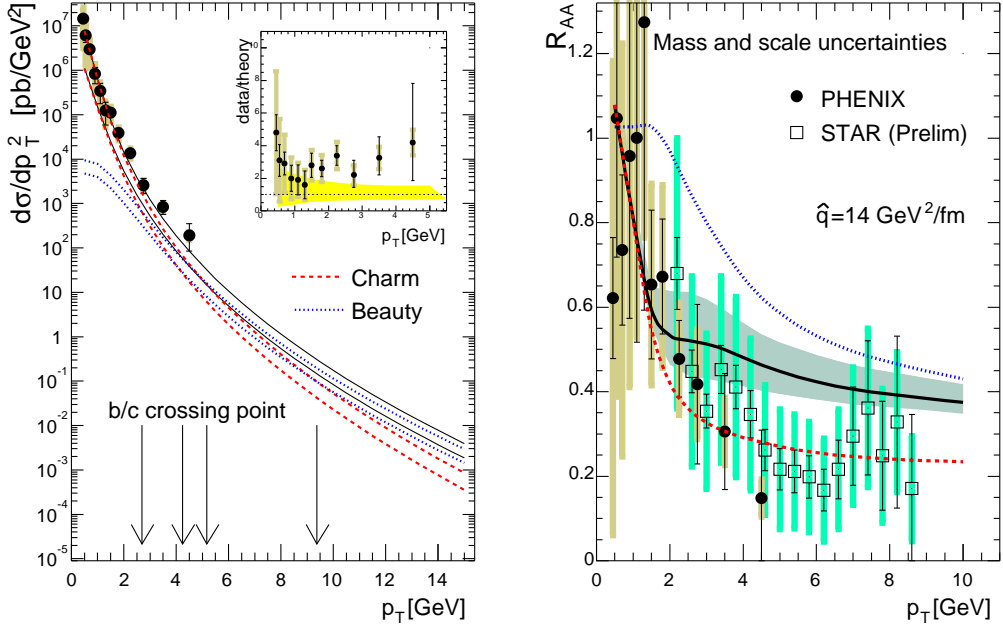}
\end{center}
\caption{(Color online) (a) Left: Comparison of the FONLL calculation of 
single inclusive electrons~\cite{Cacciari:2005rk} to data from pp collisions at 
$\sqrt{s} = 200$~GeV~\cite{Adler:2005fy}. 
Upper and lower lines are estimates of theoretical uncertainties, obtained by varying 
scales and masses, for details see text.
(b) Right: The nuclear modification factor $R_{AA}^e$ of electrons in 
central Au--Au collisions for an opacity of the produced QCD matter characterized by
the time-averaged BDMPS transport coefficient $\hat{q} = 14\, {\rm GeV}^2/{\rm fm}$.
The shaded band indicates the theoretical uncertainty of the
perturbative baseline only. Red dashed and blue dotted curves show $R_{AA}^e$
for $c$-quark and $b$-quark decay contributions, respectively. Data taken from Ref.~
\cite{Bielcik:2005wu,Adler:2005xv}.
}\label{fig2}
\end{figure}

In Fig.~\ref{fig2}a, we compare the perturbative FONLL result for
the single electron spectrum in $\sqrt{s}=200$~GeV pp
collisions~\cite{Cacciari:2005rk} to data from the PHENIX collaboration at RHIC~\cite{Adler:2005fy}. 
Upper and lower lines of the different
decay contributions indicate theoretical uncertainties: they were estimated 
by varying the factorization and renormalization scales independently in the
range $m_T/2 < \mu_{F,R} < 2\, m_T$ with the constraint $0.5<\mu_F/\mu_R<2$.
In addition, we have varied quark masses around their central values $m_c = 1.5$~GeV 
and $m_b = 4.75$~GeV over the range $1.3 < m_c < 1.7$~GeV, $4.5 < m_b < 5.0$~GeV.
These effects were then added in quadrature.
The uncertainties related to the fragmentation picture, intrinsically
ambiguous in the $p_T \lsim m$ region, were not explicitly considered here,
as they are not expected to be very large (see for instance~\cite{Cacciari:2003uh}).
Within errors, the comparison between FONLL predictions and experiment is fair,
see Fig.~\ref{fig2}a. However, 
the central value of the FONLL calculation under-estimates the central value of the measured 
electron yields by a factor 2--3. 
Thus, an ad hoc increase of the 
charm cross section by a factor of three, large but still at the borderline of the above-mentioned uncertainties, 
would bring the central values of experiment and theory 
into agreement (data not shown). This illustrates that there is only limited control over
the proton--proton baseline, on top of which one 
aims at establishing medium effects. Another illustration of this is 
that the $b$-quark decay contribution may start
dominating over the $c$ quark one at a transverse momentum as low as $p_T^e = 2.5$
GeV or as high as $p_T^e = 9.5$~GeV (see Fig.~\ref{fig2}).  This translates into a 
significant uncertainty in $R_{AA}^e$, as we discuss now.

\vspace{0.5cm}
%

\section{The  nuclear modification factor for single electrons in Au--Au collisions at RHIC}

The medium-induced suppression of high-$p_T$ electrons in nucleus--nucleus 
collisions can be characterized by the nuclear modification factor $R^e_{AA}(p_T)$, which 
compares the production of electrons in AA collisions 
to an equivalent number of pp collisions, 
\begin{equation}
R^e_{AA}(p_T)={\left.{\d^2 N^{AA\to e}_{\rm medium}/ \d p_T\,\d y}\right|_{y=0} \over
\langle N^{AA}_{\rm coll}\rangle\left.{\d^2 N^{pp\to e}/  \d p_T\,\d y}\right|_{y=0}}\, .
\label{2.1}
\end{equation}
Here,  $\langle N^{AA}_{\rm coll}\rangle$ is the average number of inelastic nucleon--nucleon 
collisions in a given centrality class. We calculate $R_{AA}^e$ according to the model
used successfully for the description of suppressed high-$p_T$ light-flavored hadron 
production~\cite{Eskola:2004cr,Dainese:2004te}. To calculate 
${\d^2 N^{AA\to e}_{\rm medium}/ \d p_T\,\d y}$, we start from the FONLL spectrum 
of final state heavy quarks in pp collisions. The nuclear modification of  parton distribution 
functions is expected to be at most a 10\% effect and can be 
safely neglected in our calculation~\cite{Armesto:2005iq}.
We calculate the medium-induced 
energy degradation of the partonic spectrum by means of quenching weights for massive
quarks~\cite{Armesto:2003jh,Armesto:2005iq}, before fragmenting the energy-degraded 
partons in the vacuum according to FONLL fragmentation functions. This assumes that the interplay
between quenching and hadronization can be neglected.
 Our treatment of medium effects is 
based on a realistic description of the collision geometry used to compute the
in-medium path length and medium density on the parton-by-parton level~\cite{Dainese:2004te,Eskola:2004cr}.

\vskip 0.2cm
\begin{figure}[htb]
\begin{center}
\includegraphics[width=8.3cm,angle=-90]{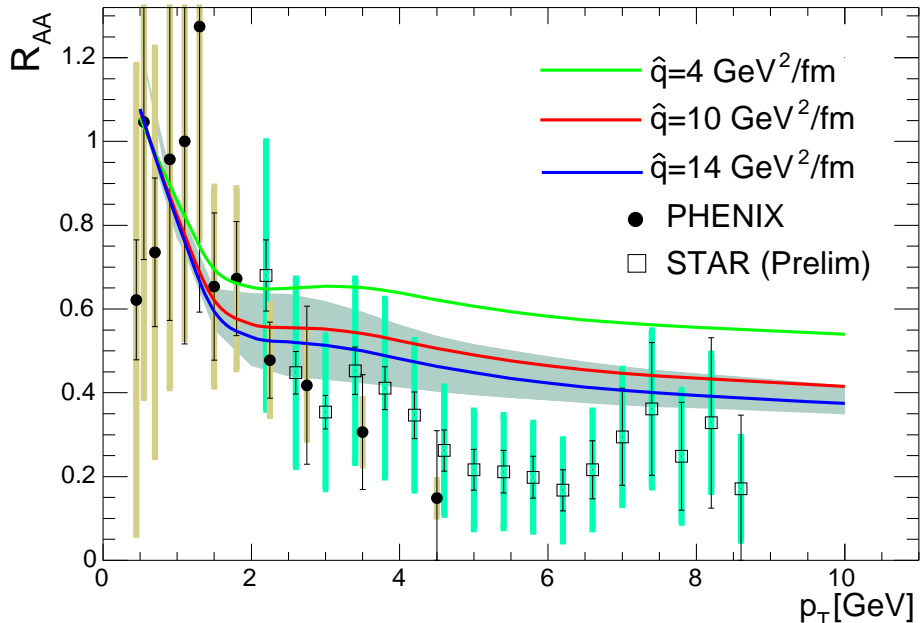}
\end{center}
\caption{(Color online)
The nuclear modification factor for single electrons in central Au--Au collisions
at RHIC. Curves indicate the suppression for different opacities of the produced matter.
The shaded band indicates the theoretical uncertainty of the
perturbative baseline for $\hat{q}=14\, {\rm GeV}^2/{\rm fm}$.
}
\label{fig3}
\end{figure}
%

The effect of parton
energy loss on single inclusive spectra depends on the color charge and mass of 
the parent parton, its in-medium path length $L$, and the time-averaged squared
momentum transfer from the medium to the partonic projectile, which is  characterized 
by the BDMPS transport coefficient $\hat{q}$~\cite{Baier:1996sk}.
Previous studies have shown that for $4< \hat{q} < 14\, {\rm GeV}^2/{\rm fm}$, this 
model of parton energy loss accounts for the strength and approximate $p_T$-independence 
of light hadron suppression at RHIC~\cite{Dainese:2004te,Eskola:2004cr}. 
In Fig.~\ref{fig2}b, we show the nuclear modification factor (\ref{2.1}) for
$\hat{q} = 14\, {\rm GeV}^2/{\rm fm}$. 
For this large value of $\hat{q}$, a calculation anchored on the FONLL perturbative 
baseline tends to over-estimate $R_{AA}^e$. However, due to the sizable theoretical and 
experimental 
errors,  claims about an inconsistency between theory and experiment are not supported 
by Fig.~\ref{fig2}b. We observe that an experimental separation of $b$- and $c$-
decay electrons would strongly enhance the sensitivity to the mass hierarchy of parton
energy loss. Indeed, since the mass dependence of charm quark energy loss vanishes as 
a function of $(m_c/p_T^c)^2$, the suppression of electrons from charm decays is comparable 
to that of light-flavored hadrons. In contrast, electrons from $b$-decays are much less suppressed
due to the larger mass of their quark parents, see Fig.~\ref{fig2}b and 
Ref.~\cite{Djordjevic:2005db}. 

In Fig.~\ref{fig3}, we compare the same data on the nuclear modification factor to 
central values of the parton energy loss model, calculated for different values of 
the transport coefficient $\hat{q}$. By comparing  Fig.~\ref{fig3} with the corresponding
results for the nuclear modification factor $R_{AA}^h$ for light hadrons~\cite{Dainese:2004te,Eskola:2004cr},
we find that for the experimentally favored range $\hat{q}=4-14\, {\rm GeV}^2/{\rm fm}$, 
the difference between $R_{AA}^e$ and $R_{AA}^h$ is roughly 0.2.
In general, the theoretical 
uncertainty in calculating the partonic baseline spectrum is comparable to the model-intrinsic 
uncertainty in determining $\hat{q}$. For instance,  the central value for 
$\hat{q} = 10\, {\rm GeV}^2/{\rm fm}$ lies within the perturbative uncertainty band 
of $\hat{q} = 14\, {\rm GeV}^2/{\rm fm}$. We also note that as $\hat{q}$ increases, the 
suppression gradually saturates. This is due to the fact that at 
high opacity, particle production becomes surface biased, thereby decreasing the 
sensitivity of the nuclear modification factor on $\hat{q}$~\cite{Eskola:2004cr,Muller:2002fa}. 
Since heavy quarks lose less energy 
than light quarks or gluons, this surface bias becomes important at higher values 
of $\hat{q}$. Thus, if experimental
and theoretical errors can be controlled, the measurement of the decay products of
heavy quarks does have the potential to improve the characterization of the very
opaque medium observed at RHIC~\cite{Djordjevic:2005db}.

\begin{figure}[htb]
\begin{center}
\includegraphics[width=10.3cm]{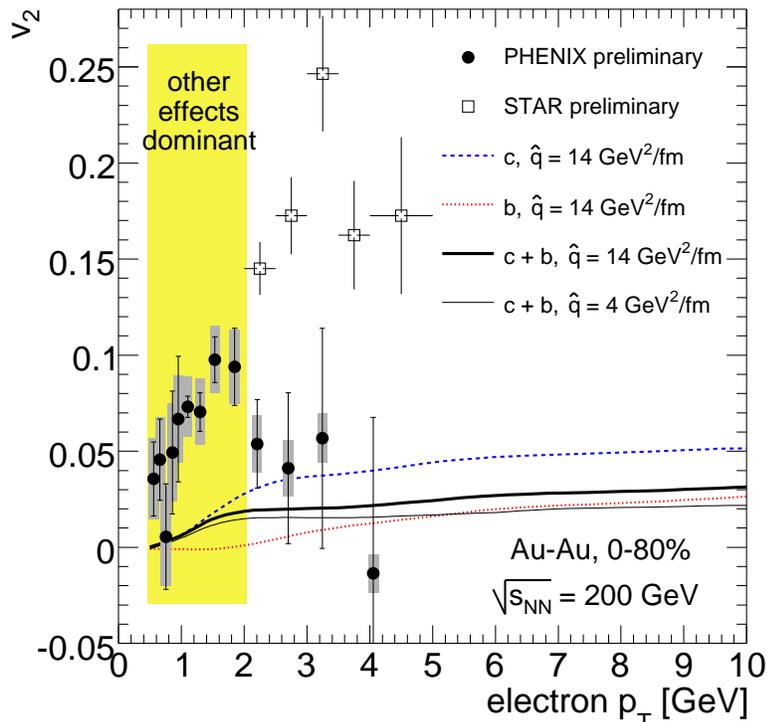}
\end{center}
\caption{(Color online) The azimuthal asymmetry of the single inclusive electron spectrum 
in Au--Au collisions at 0-80 \% centrality, compared to our model of parton energy loss. Data
taken from Refs.~\cite{Sakai:2005qn,Laue:2004tf}.
 }
\label{fig4}
\end{figure}

Calculating in `minimum bias' (0--80 \%) Au--Au collisions the angular 
dependence of the yield of heavy quark
decay electrons with respect to the reaction plane, we can determine
the azimuthal asymmetry, expected from parton energy loss in a spatially
asymmetric medium~\cite{Wang:2000fq,Armesto:2004vz}. 
This asymmetry is characterized by the so-called elliptic flow,
which is the second harmonic coefficient $v_2^e$ in the Fourier decomposition of
the angular dependence $\phi$ with respect to the orientation $\Psi_R$ of the reaction
plane, $dN^{AA \to e}/p_T\, dp_T\, d\phi \propto 1 + 2\, v_2^e(p_T)\, \cos\left[2(\phi - \Psi_R
)\right]$~\cite{Borghini:2000sa}.
At low transverse momenta, 
other collective phenomena are expected to dominate $v_2^e$ for heavy quarks and 
their decay electrons~\cite{Lin:2003jy,Greco:2003vf,Moore:2004tg,vanHees:2004gq,vanHees:2005wb,Zhang:2005ni}.
However, for $p_T^e > 2$~GeV, the azimuthal dependence of the energy loss of 
heavy quarks may be the dominant contribution to the observed asymmetry. As seen in
Fig.~\ref{fig4}, preliminary data in this kinematic regime show significant uncertainties, and
our calculation favors a non-vanishing but small value for the high-$p_T$ azimuthal asymmetry of
electrons. In the presence of collective transverse flow effects, the BDMPS transport
coefficient may be sensitive to a combination of energy density and directed collective
motion. This is expected to further increase high-$p_T$  $v_2^e$~\cite{Armesto:2004vz,Renk:2005ta}. 
Thus, our present study does not
allow us to exclude values of $v_2^e$ which are somewhat larger than $v_2^e = 0.05$,
but it strongly disfavors $v_2^e > 0.1$. 
%

\section{Conclusions}
Our study indicates that for the parameter range $\hat{q}=4-14\, {\rm GeV}^2/{\rm fm}$ of
the BDMPS transport coefficient favored by the suppression of light hadron spectra at
RHIC, the nuclear modification factor $R_{AA}^e$ of single electrons  is $\sim 0.2$ 
larger than that of light-flavored hadrons. 
Thus, single electron spectra are in principle sensitive to the expected mass hierarchy of parton energy 
loss. Moreover, since massive quarks lose less energy than light ones in the medium, their 
spectrum starts being dominated by surface emission for higher medium densities only. 
Thus, single electron spectra may increase the experimental 
sensitivity to the properties of the produced matter in the region of high opacity.

However, to exploit these opportunities of single electron measurements in nucleus--nucleus
collisions demands a significantly better control of the various sources of uncertainties. The
preliminary data available so far neither allow us to support claims of an inconsistency between
theory and experiment, nor do they support yet the theoretically expected mass hierarchy
of nuclear modification factors. Here, we have contributed to the generally needed assessment
of uncertainties by characterizing the theoretical ones. In particular, we have established
that uncertainties in the  perturbative baseline are significant, and should not  be neglected in calculations of the nuclear modification factor. 
If future experiments succeeded in disentangling the $b$- and $c$-decay contributions
(e.g. by reconstructing the displaced decay vertices), this would remove a significant source 
of theoretical uncertainty, dramatically enhancing the experimental sensitivity to the mass hierarchy 
of parton energy loss, as can be seen from Fig.~\ref{fig2}b. As an aside, we note that 
single electron spectra are not the only possibility to characterize the mass dependence of
parton energy loss. Decay muons may provide another, as yet unexplored possibility. 
Also, topologically reconstructed $D$- and $B$-mesons provide an alternative and much 
more versatile access to the same information, which is particularly promising at the 
LHC~\cite{Armesto:2005iq,Dainese:2005me}. 

We finally emphasize that, although preliminary RHIC data favor high values 
($\hat{q}=14\, {\rm GeV}^2/{\rm fm}$) of the BDMPS transport coefficient,
significantly smaller densities given by  $\hat{q}=10\, {\rm GeV}^2/{\rm fm}$ reproduce
the data almost as satisfactorily, since their theoretical uncertainties largely overlap (see Fig.~\ref{fig3}). 
The densities required in our calculation to obtain $R_{AA}^e(p_T^e=10\, {\rm GeV}) = 0.4$
are not larger than those used previously~\cite{Dainese:2004te,Eskola:2004cr} 
for the characterization of light hadron suppression.

\noindent {\bf Acknowledgements}\\
We thank Paolo Nason for providing a fast code for performing the fragmentation convolution 
and the decay of the heavy quarks.
We also thank John Campbell and Fabio Maltoni for discussions about the 
calculation of the Drell--Yan spectrum at NLO.
N.A. acknowledges financial support of Ministerio de Educaci\'on y Ciencia
of Spain under a contract Ram\'on y Cajal, and of CICYT of Spain under
project FPA2002-01161.



\begin{thebibliography}{99}
%
\bibitem{Dokshitzer:2001zm}
Y.~L.~Dokshitzer and D.~E.~Kharzeev,
Phys.\ Lett.\ B {\bf 519}  (2001) 199.
%
\bibitem{Armesto:2003jh}
N.~Armesto, C.~A.~Salgado and U.~A.~Wiedemann,
Phys. Rev. D {\bf 69} (2004) 114003.
%
\bibitem{Zhang:2003wk}
B.~W.~Zhang, E.~Wang and X.~N.~Wang,
Phys.\ Rev.\ Lett.\  {\bf 93} (2004) 072301.
%
\bibitem{Djordjevic:2003zk}
M.~Djordjevic and M.~Gyulassy,
Nucl.\ Phys.\ A {\bf 733} (2004) 265.

%
\bibitem{Wang:2003aw}
X.~N.~Wang,
Phys.\ Lett.\ B {\bf 579} (2004) 299.
%
\bibitem{Dainese:2004te}
A.~Dainese, C.~Loizides and G.~Paic,
Eur.\ Phys.\ J.\ C {\bf 38} (2005) 461.
%
\bibitem{Eskola:2004cr}
K.~J.~Eskola, H.~Honkanen, C.~A.~Salgado and U.~A.~Wiedemann,
Nucl.\ Phys.\ A {\bf 747} (2005) 511.
%
\bibitem{Gyulassy:2003mc}
  M.~Gyulassy, I.~Vitev, X.~N.~Wang and B.~W.~Zhang,
  arXiv:nucl-th/0302077.
 %
\bibitem{Jacobs:2004qv}
  P.~Jacobs and X.~N.~Wang,
  Prog.\ Part.\ Nucl.\ Phys.\  {\bf 54} (2005) 443.

%
\bibitem{Adcox:2004mh}
  K.~Adcox {\it et al.}  [PHENIX Collaboration],
  Nucl.\ Phys.\ A {\bf 757} (2005) 184.
%
\bibitem{Back:2004je}
  B.~B.~Back {\it et al.} [PHOBOS Collaboration],
  Nucl.\ Phys.\ A {\bf 757} (2005) 28.
%
\bibitem{Arsene:2004fa}
  I.~Arsene {\it et al.}  [BRAHMS Collaboration],
  Nucl.\ Phys.\ A {\bf 757} (2005) 1.
%
\bibitem{Adams:2005dq}
  J.~Adams {\it et al.}  [STAR Collaboration],
  Nucl.\ Phys.\ A {\bf 757} (2005) 102.

\bibitem{Armesto:2005iq}
  N.~Armesto, A.~Dainese, C.~A.~Salgado and U.~A.~Wiedemann,
  Phys.\ Rev.\ D {\bf 71} (2005) 054027.

\bibitem{Bielcik:2005wu}
  J.~Bielcik [STAR Collaboration],
  arXiv:nucl-ex/0511005.
  
\bibitem{Adler:2005xv}
  S.~S.~Adler  {\it et al.} [PHENIX Collaboration],
  arXiv:nucl-ex/0510047.
 
\bibitem{Adcox:2002cg}
  K.~Adcox {\it et al.}  [PHENIX Collaboration],
  Phys.\ Rev.\ Lett.\  {\bf 88} (2002) 192303.
  
\bibitem{Sakai:2005qn}
  S.~Sakai [PHENIX Collaboration],
  arXiv:nucl-ex/0510027.

\bibitem{Laue:2004tf}
  F.~Laue  [STAR Collaboration],
  J.\ Phys.\ G {\bf 31} (2005) S27; F.~Laue, talk at Quark Matter 2005.

\bibitem{Nason:1987xz}
P.~Nason, S.~Dawson and R.~K.~Ellis,
Nucl.\ Phys.\ B {\bf 303} (1988) 607.

\bibitem{Nason:1989zy}
P.~Nason, S.~Dawson and R.~K.~Ellis,
Nucl.\ Phys.\ B {\bf 327} (1989) 49
[Erratum-ibid.\ B {\bf 335} (1990) 260].

\bibitem{Beenakker:1990ma}
W.~Beenakker, W.~L.~van Neerven, R.~Meng, G.~A.~Schuler and J.~Smith,
Nucl.\ Phys.\ B {\bf 351} (1991) 507.

\bibitem{Cacciari:2005rk}
  M.~Cacciari, P.~Nason and R.~Vogt,
  Phys.\ Rev.\ Lett.\  {\bf 95} (2005) 122001.

\bibitem{Cacciari:1998it}
  M.~Cacciari, M.~Greco and P.~Nason,
  JHEP {\bf 9805} (1998) 007.
  
\bibitem{Cacciari:2001td}
  M.~Cacciari, S.~Frixione and P.~Nason,
  JHEP {\bf 0103} (2001) 006.

\bibitem{Sjostrand:2001yu}
T.~Sjostrand, L.~Lonnblad and S.~Mrenna,
arXiv:hep-ph/0108264.

\bibitem{Gavin:1995ch}
  S.~Gavin, R.~Kauffman, S.~Gupta, P.~V.~Ruuskanen, D.~K.~Srivastava and
R.~L.~Thews,
  Int.\ J.\ Mod.\ Phys.\ A {\bf 10} (1995) 2961.

\bibitem{Adler:2005fy}
  S.~S.~Adler {\it et al.}  [PHENIX Collaboration],
  arXiv:hep-ex/0508034.

\bibitem{Cacciari:2003uh}
   M.~Cacciari, S.~Frixione, M.~L.~Mangano, P.~Nason and G.~Ridolfi,
   JHEP {\bf 0407} (2004) 033.

\bibitem{Baier:1996sk}
  R.~Baier, Y.~L.~Dokshitzer, A.~H.~Mueller, S.~Peigne and D.~Schiff,
  Nucl.\ Phys.\ B {\bf 484} (1997) 265.

\bibitem{Djordjevic:2005db}
  M.~Djordjevic, M.~Gyulassy, R.~Vogt and S.~Wicks,
  arXiv:nucl-th/0507019.

\bibitem{Muller:2002fa}
  B.~Muller,
  Phys.\ Rev.\ C {\bf 67} (2003) 061901.

\bibitem{Wang:2000fq}
  X.~N.~Wang,
  Phys.\ Rev.\ C {\bf 63} (2001) 054902.
  
\bibitem{Armesto:2004vz}
N.~Armesto, C.~A.~Salgado and U.~A.~Wiedemann,
arXiv:hep-ph/0411341.

\bibitem{Borghini:2000sa}
  N.~Borghini, P.~M.~Dinh and J.~Y.~Ollitrault,
  Phys.\ Rev.\ C {\bf 63} (2001) 054906.

\bibitem{Lin:2003jy}
  Z.~w.~Lin and D.~Molnar,
  Phys.\ Rev.\ C {\bf 68} (2003) 044901.

\bibitem{Greco:2003vf}
  V.~Greco, C.~M.~Ko and R.~Rapp,
  Phys.\ Lett.\ B {\bf 595} (2004) 202.

\bibitem{Moore:2004tg}
  G.~D.~Moore and D.~Teaney,
  Phys.\ Rev.\ C {\bf 71} (2005) 064904.

\bibitem{vanHees:2004gq}
  H.~van Hees and R.~Rapp,
  Phys.\ Rev.\ C {\bf 71} (2005) 034907.

\bibitem{vanHees:2005wb}
  H.~van Hees, V.~Greco and R.~Rapp,
  arXiv:nucl-th/0508055.
   
\bibitem{Zhang:2005ni}
  B.~Zhang, L.~W.~Chen and C.~M.~Ko,
  Phys.\ Rev.\ C {\bf 72} (2005) 024906.


\bibitem{Renk:2005ta}
  T.~Renk and J.~Ruppert,
  Phys.\ Rev.\ C {\bf 72} (2005) 044901.

\bibitem{Dainese:2005me}
  A.~Dainese  [ALICE Collaboration],
  arXiv:nucl-ex/0510082; 
  ALICE Physis Performance Report, Volume II, CERN/LHCC 2005-030 (2005).







\end{thebibliography}
\end{document}